\documentclass[shortnote]{jpsj2} 
\title{High rate production of polarized ${}^{3}{\rm He}$ with 
meta-stability exchange method}

\author{
Ema \textsc{Ihara}$^{1}$
\thanks{E-mail address: ihara@phys.kyushu-u.ac.jp},
Tomotsugu \textsc{Wakasa}$^{1}$
\thanks{E-mail address: wakasa@phys.kyushu-u.ac.jp},
Masanori \textsc{Dozono}$^{1}$, and
Yasuhiro \textsc{Sakemi}$^{2}$}

\inst{
$^{1}$Department of Physics, Kyushu University, 
Fukuoka 812-8581 \\
$^{2}$Cyclotron and Radioisotope Center, Tohoku University, 
Miyagi 980-8578}

\kword{
polarized ${}^{3}{\rm He}$, 
meta-stability exchange,
infrared laser}

\begin{document}
\maketitle

 Recent progress in the development of high intensity 
infrared fiber lasers enables us to produce highly polarized 
${}^{3}{\rm He}$ nuclei by the meta-stability exchange method
\cite{pr_132_2561_1963}.
 Polarization with $P\gtrsim 0.8$ has been realized 
\cite{jp_2_2159_1992,nima_320_53_1992,pra_47_456_1993}, 
and the polarized ${}^{3}{\rm He}$ gas 
has been used as polarized ${}^{3}{\rm He}$ targets 
for nuclear physics experiments as well as 
signal source for lung magnetic resonance imaging (MRI).
 The meta-stability exchange method produces polarized 
${}^{3}{\rm He}$ nuclei much faster than the 
spin-exchange method because the pumping rate, i.e.,
the rate at which ${}^{3}{\rm He}$ nuclei can be polarized, 
is much higher.
 This advantage is suitable for applications such as 
on-demand production for lung MRI.
 Another important feature of the meta-stability exchange method 
is simultaneous production of polarized ${}^{3}{\rm He}^+$
ions.
 The electrons of ${}^{3}{\rm He}^+$ ions are also polarized, and 
thus they can be applied for investigation of material surfaces
\cite{priv_hidaka}.
 For polarized ${}^{3}{\rm He}$ targets, higher nuclear 
polarization has been required for experiments, and 
higher polarization can be achieved with slower
relaxation rate resulting in slower pumping rate.
 For the above-mentioned applications, on the contrary, 
faster pumping rate with reasonable nuclear 
polarization is suitable.
 Therefore, we have investigated the relation between 
the pumping rate and the nuclear polarization in high 
pumping rate region.

\begin{figure}[p]
\begin{center}
\includegraphics[width=0.8\textwidth]{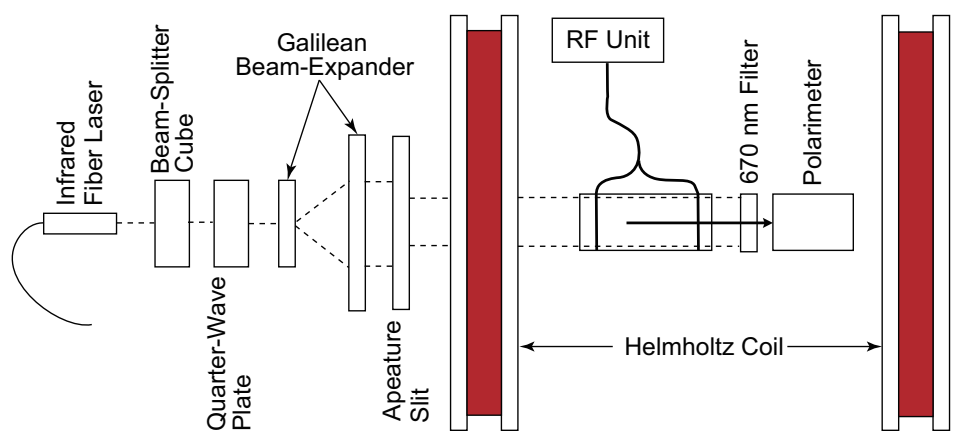}
\end{center}
\caption{
 Schematic view of the PLUM system.
}
\label{fig:setup}
\end{figure}

 The meta-stability exchange method is an extension 
of ordinary optical pumping method \cite{rmp_44_169_1972}.
 Figure~\ref{fig:setup} shows a schematic of our PLUM system
(PLUM stands for Polarizer with Laser Using Meta-stability 
exchange).
 A ${}^{3}{\rm He}$ gas with a pressure of 
0.3 Torr is sealed in a Pyrex glass cell 
with a size of $3\,{\rm cm}^\phi\times 5\, {\rm cm}^t$.
 This ${}^{3}{\rm He}$ cell is set in a uniform magnetic 
field of 13 G generated by a Helmholtz coil in 
order to keep the polarization.
 In the meta-stability exchange method, first
${}^{3}{\rm He}$ atoms are excited to the meta-stable
$2{}^{3}{\rm S}_1$ state by applying an RF field 
with a frequency of $f$=1--10 MHz.
 Secondly, ${}^{3}{\rm He}$ atoms in $2{}^{3}{\rm S}_1$ 
are optically pumped to the $2{}^{3}{\rm P}$ state 
with $\sim$1083 nm infrared light produced by a 
Keopsys fiber laser module \cite{keopsys}.
 We use a linearly polarizing beam-splitter cube followed by
a quarter-wave plate to circularly polarize the laser light.
 For the cell with 3 cm diameter, we expand the beam 
with a Galilean type beam expander.
 By using left-handed circularly polarized light,
only two sublevels of $M_F$ = $-3/2$ and $-1/2$ out of
four sublevels of $M_F$ = $\pm 3/2$ and $\pm 1/2$ in
$2{}^{3}{\rm S}_1$ are concerned with the optical pumping,
which results in atomic (= total spin $F$) polarization 
for ${}^{3}{\rm He}$ atoms in the meta-stable 
$2{}^{3}{\rm S}_1$ state.
 There are nine transitions $C_1$--$C_9$ 
between $2{}^{3}{\rm S}_1$ and $2{}^{3}{\rm P}$, and 
$C_8$ and $C_9$ transitions are known to be efficient 
for production of highly polarized ${}^{3}{\rm He}$
\cite{jp_46_2057_1985}.
 The $C_8$ and $C_9$ transitions correspond to 
($2{}^{3}{\rm S}_1$,$F=1/2$)
$\rightarrow$
($2{}^{3}{\rm P}_0$,$F=1/2$) and 
($2{}^{3}{\rm S}_1$,$F=3/2$)
$\rightarrow$
($2{}^{3}{\rm P}_0$,$F=1/2$), respectively.
 Finally, the atomic polarization is transferred to 
the nuclear polarization of the ground state by
meta-stability exchange collisions.

\begin{figure}[p]
\begin{center}
\includegraphics[width=0.8\textwidth]{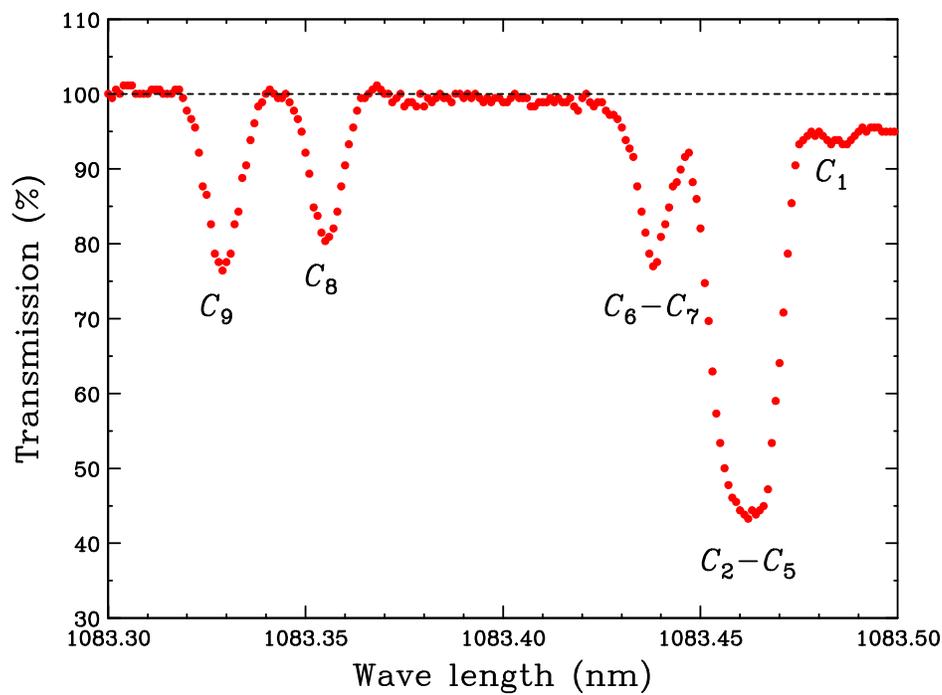}
\end{center}
\caption{
 Transmission of infrared light 
in ${}^{3}{\rm He}$ gas as a 
function of wave length.
 The $C_1$--$C_9$ peaks correspond to the resonance 
absorptions from $2{}^{3}{\rm S}_1$ to 
$2{}^{3}{\rm P}$.
}
\label{fig:att}
\end{figure}

 Figure~\ref{fig:att} shows the measured transmission of
infrared light as a function of wave length 
with an RF discharge frequency of $f$ = 6.5 MHz.
 Resonance absorptions for $C_1$--$C_9$ transitions 
are clearly observed in the spectrum.
 The observed width of $\sim$2 GHz corresponds to the 
intrinsic linewidth of the fiber laser, which 
is suitable for efficient optical pumping because the 
linewidth well matches to the $\sim 2$ GHz Doppler bandwidth 
of the ${}^{3}{\rm He}$ gas \cite{pra_47_456_1993}.
 This narrow linewidth is sufficient to resolve 
the 6.7 GHz hyperfine splitting in the $2{}^{3}{\rm S}_1$ 
state, which enables us to use the $C_8$ and $C_9$ transitions
separately.

\begin{figure}[p]
\begin{center}
\includegraphics[width=0.8\textwidth]{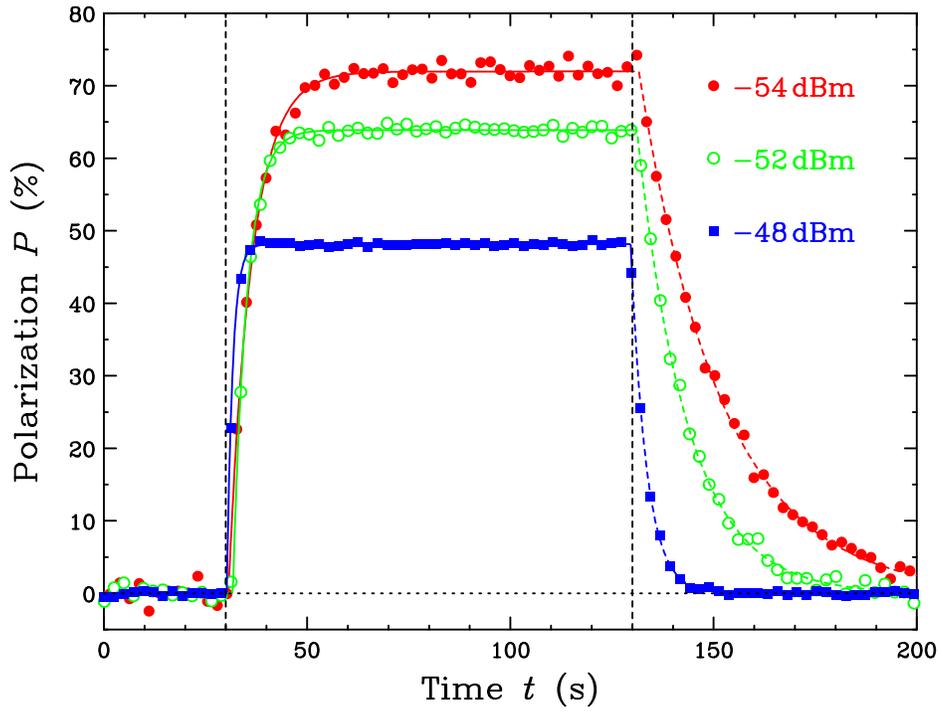}
\end{center}
\caption{
 Build-up and relaxation of nuclear polarization $P$ 
as a function of time. 
 See text for details.}
\label{fig:build_relax}
\end{figure}

 The nuclear polarization of ${}^{3}{\rm He}$ can be obtained 
by measuring the circular polarization of an optical line 
at 668 nm ($3{}^{1}{\rm D}_2$ $\rightarrow$ $2{}^{1}{\rm P}_1$)
\cite{jp_2_2159_1992,pra_47_468_1993}.
 An isolation of 668 nm light is performed using a 
Thorlabs laserline filter FL670 \cite{thorlabs}.
 The circular polarization of the isolated light 
is measured using a Thorlabs polarization analyzing 
system PAX5710VIS \cite{thorlabs}.
 Figure~\ref{fig:build_relax} shows 
the typical ${}^{3}{\rm He}$ nuclear polarization 
deduced from the circular polarization of 668 nm light 
as a function of time.
 The laser was tuned for the $C_8$ transition, and 
was irradiated from $t$ = 30 to 130 s 
with an RF discharge frequency of $f$ = 8.3 MHz.
 The measurements were performed for several RF discharge 
intensities which resulted in 668 nm light powers of 
$-54$ $\sim$ $-48$ dBm on the system.
 The nuclear polarization $P$ reaches its saturation value 
with an effective laser power of $\sim$400 mW on the 
cell, and it is insensitive to the applied RF frequencies. 
 The time dependence of $P$ is expressed \cite{jp_46_2057_1985} 
as $P_0[1-\exp(-t/\tau_p)]$ 
where $P_0$ is the final polarization for $t$ $\rightarrow$ $\infty$ 
and $\tau_p$ is the pumping time, and the solid curves in 
Fig.~\ref{fig:build_relax} are the results of fitting.
 The pumping time $\tau_p$ was short as 1--6 s, 
which is an unique feature of the meta-stability exchange method.
 The maximum nuclear polarization of $P_0$ = 72\% 
was obtained in $-54$ dBm case.
 The relaxation of $P$ after stopping the laser irradiation 
is expressed \cite{jp_46_2057_1985} 
as $P_0[\exp(-t/\tau_r)]$ where $\tau_r$ is the 
relaxation time, and the dashed curves in 
Fig.~\ref{fig:build_relax} are the results of fitting.
 The relaxation time $\tau_r$ was long as 3--19 s compared with the 
pumping time $\tau_p$.

\begin{figure}[p]
\begin{center}
\includegraphics[width=0.8\textwidth]{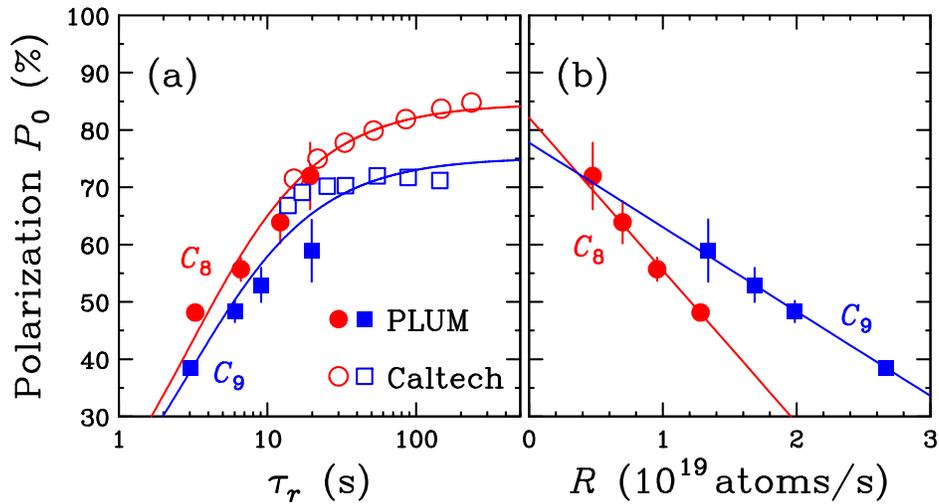}
\end{center}
\caption{
 (a) Nuclear polarization $P_0$ as a function of 
relaxation time $\tau_r$ for $C_8$ and $C_9$ transitions.
 The data in large $\tau_r$ region are the Caltech data
\cite{pra_47_456_1993}.
 The solid curves are the results of fitting with Eq.~(\ref{eq:p0}).
 (b) Nuclear polarization $P_0$ as a function of 
pumping rate $R$ for $C_8$ and $C_9$ transitions.
 The solid lines are the results of fitting with linear functions.
}
\label{fig:pol}
\end{figure}

 The relation between $\tau_r$ and $P_0$ is shown in 
Fig.~\ref{fig:pol}(a).
 The Caltech data \cite{pra_47_456_1993} 
for a 0.3 Torr cell with $f$ = 10 MHz 
are also represented in the large $\tau_r$ region.
 It is found that in the whole $\tau_r$ region,
the $C_8$ transition is the better choice to obtain 
higher polarization $P_0$ at fixed $\tau_p$.
 The nuclear polarization $P_0$ can be expressed \cite{jp_46_2057_1985} 
using $\tau_p$ and $\tau_r$ as
\begin{equation}
P_0 = P_{\infty}\frac{1}{1+\tau_p/\tau_r}\ ,
\label{eq:p0}
\end{equation}
where $P_{\infty}$ is the maximum polarization for $\tau_r$
$\rightarrow$ $\infty$.
 The solid curves in Fig.~\ref{fig:pol} are the results 
of fitting with Eq.~(\ref{eq:p0}) with constant $\tau_p$, 
which reproduce the measured data reasonably well.

 The pumping rate $R$ can be defined 
\cite{pra_47_456_1993} as $R=NP_0/\tau_p$
where $N$ is the number of ${}^{3}{\rm He}$ atoms in the cell. 
 The relations between $R$ and $P_0$ are displayed 
in Fig.~\ref{fig:pol}(b) for the measurements with $f$ = 9.6 MHz.
 The nuclear polarization $P_0$ decreases almost linearly as the 
pumping rate $R$ increases.
 The solid curves are reproduction of the data with 
linear functions, which reproduce the 
data very well.
 In the low $R$ region of $R\lesssim 4\times 10^{18}\,\mathrm{atoms/s}$, 
the $C_8$ transition is the 
better choice to obtain higher polarization $P_0$, 
and this transition has been generally used to prepare 
polarized ${}^{3}{\rm He}$ targets.
 In the present high $R$ region, on the contrary, 
the $C_9$ transition is the better choice because 
the nuclear polarization
less depends on the pumping rate and thus it takes 
a higher value in this region.
 This is mainly due to a smaller $\tau_p$ value for $C_9$ 
compared with the corresponding $\tau_p$ for $C_8$, 
and thus the pumping rate $R$ for $C_9$ becomes higher.
 It is found that high rate production of $R\simeq 2\times 10^{19}\,$ 
is possible with keeping the polarization $P_0\simeq$50\% 
by using the $C_9$ transition.
 This high rate is useful for applications which require
on-demand production such as lung MRI.

%
%
 We are grateful to P. J. Nacher and G. Tastevin at ENS,
E. W. Otten, W. Heil, R. K. Kremer and M. Batz at Maintz, and 
H. Sasada at Keio for our interaction and their 
valuable suggestions.
 This work was supported in part by the 
Grant-in-Aid for Scientific Research No. 
16654064 
of the Ministry of Education, Culture, Sports, 
Science, and Technology of Japan.

%
%

%
%


\begin{thebibliography}{99} 
\bibitem{pr_132_2561_1963}
F. D. Colegrove, L. D. Chaearer and G. K. Walters:
Phys. Rev. \textbf{132} (1963) 2561.
\bibitem{jp_2_2159_1992}
N. P. Bigelow, P. J. Nacher and M. Leduc:
J. Phys. II France \textbf{2} (1992) 2159.
\bibitem{nima_320_53_1992}
G. Eckert et al.:
Nucl. Instrum. Methods A \textbf{320} (1992) 53.
\bibitem{pra_47_456_1993}
T. R. Gentile and R. D. McKeown:
Phys. Rev. A \textbf{47} (1993) 456.
\bibitem{priv_hidaka}
M. Hidaka: private communication.
\bibitem{rmp_44_169_1972}
W. Happer,
Rev. Mod. Phys. \textbf{44} (1972) 169.
\bibitem{keopsys}
http://www.keopsys.com
\bibitem{pra_47_468_1993}
W. Lorenzon, T. R. Gentile, H. Gao and R. D. McKeown:
Phys. Rev. A \textbf{47} (1993) 468.
\bibitem{thorlabs}
http://www.thorlabs.com
\bibitem{jp_46_2057_1985}
P. J. Nacher and M. Leduc:
J. Physique \textbf{46} (1985) 2057.

\end{thebibliography}
\end{document}